\documentclass[review]{elsarticle}

\usepackage{amsmath}

\usepackage{xcolor}

\journal{Journal of \LaTeX\ Templates}

\newcommand{\sci}[2]{\ensuremath{#1\!\times\!10^{#2}}}

\begin{document}

\begin{frontmatter}

  \title{A Method for Proton Detector Efficiency Measurement with Applications in Neutron Beta Decay and Lifetime Experiments} 

\author[lanladdress]{Grant V. Riley}
\author[utaddress]{Nadia Fomin \corref{correspondingauthor}}
\author[msuaddress]{Jaideep Taggart Singh}
\author[utaddress]{William Greene}
\author[utaddress]{Rebecca Godri}
\author[utaddress]{Evan Adamek}
\author[utaddress]{Eli Carter}
\cortext[correspondingauthor]{Corresponding Author}
\address[utaddress]{University of Tennessee, Knoxville}
\address[msuaddress]{640 S. Shaw Lane, Facility for Rare Isotope Beams, Michigan State University}
\address[lanladdress]{Los Alamos National Laboratory}

\begin{abstract} 
  This article proposes a new method to measure the proton detector efficiency for use in ``beam'' 
  determinations of the free neutron lifetime.  There is currently a $4\sigma$ disagreement 
  between the ``beam'' and ``storage'' methods of measuring the lifetime of the 
  neutron.  A possible reason for this is a systematic uncertainty that is not properly accounted for in one 
  or both types of experiments.
  Absolute proton counting is an essential facet of the ``beam'' experimental
  approach.  The absolute detector efficiency is not currently
  known for existing experiments and
  could be a source of a hidden systematic error. The proposed absolute calibration
  techniques can also be extended to new, large area particle detectors designed for future
  experiments. We propose to laser ionize a beam of atomic hydrogen and 
  simultaneously count both the outgoing protons and electrons. Due to recent advances in UV laser and atomic hydrogen beam
  technologies, we estimate a count rate as high as $\sim$200k proton-electrons pairs 
  per second under the most favorable scheme discussed. Our proposed technique 
  would remove one of the leading systematic uncertainties in the 
  determination of the neutron lifetime, making an precision of $\pm 0.16$~s 
  feasible. Precise and accurate knowledge of the proton detection efficiency is 
  crucial to fully understanding the systematics of the ``beam'' measurement 
  method as well as developing next generation experiments that will have the
  precision to test the Standard Model.
\end{abstract}

\begin{keyword}
  \texttt{neutron physics, neutron lifetime, detector efficiency}
\end{keyword}

\end{frontmatter}

\section{Motivation}

Neutron beta decay is a process that has been under investigation since the discovery of the
neutron in the early 1930s. However, even today, the free neutron lifetime is not known to a 
satisfactory level of accuracy~\cite{Wietfeldt11, Abele2008}.  
The neutron lifetime is an input to Big Bang Nucleosynthesis calculations
which determine light element abundances in the early universe~\cite{Planck, iocco09, nollett14} 
and the uncertainty on the lifetime is the most significant uncertainty in these calculations. 
The best measurements of the neutron lifetime can currently achieve 0.36~s total uncertainties~\cite{gonzalez}.
The neutron lifetime is a necessary component to determining the $V_{ud}$ element of the CKM quark 
mixing matrix in the standard model (SM) of particle physics~\cite{Hardy09, Hardy2014}. Non-unitarity of the CKM
matrix would signal new physics beyond the Standard Model. The best measurements of $V_{ud}$ currently come
from nuclear beta decay~\cite{Hardy09, Hardy2014}, and as such are subject to nuclear structure corrections. A result from neutron beta
decay would be a determination of $V_{ud}$ free from these corrections. 
This is not sufficient for tests of the Standard Model via $V_{ud}$
determinations, where a $<$0.3~s precision in necessary.
In order to reach this precision regime, great care must be taken to understand and characterize 
the systematic uncertainties involved in the measurements.  An independent and precise method 
for the evaluation of the efficiency of proton detectors used in ``beam'' determinations of the 
neutron lifetime is required to move forward.

\subsection{Neutron Beta Decay}  

Neutron beta decay is the archetype for all nuclear beta decay processes.  It makes for a very simple 
system to study the charged vector current component of the weak interaction as the neutron is a single 
nucleon, rather than a composite nuclear environment, making results easier to interpret.  
Within the Standard Model, neutron beta decay can (to a good approximation) be completely described by two 
quantities, $g_V$ and $g_A$, which give the strength of the vector and axial-vector couplings in the weak 
interaction.  Study of neutron beta decay includes several other observables, including the lifetime of the free neutron as 
well as several correlation parameters. A measurement of the lifetime probes a linear combination of 
$g_V^2$ and $g_A^2$, whereas several of the other correlation parameters give the ratio, $g_A/g_V$.  
Combining the measurements of multiple observables completely constrains the standard model parameters.

The neutron beta decay probability is given in Ref.~\cite{Jackson57} as:

\begin{equation}
\label{eq:decayrate}
dW \propto (g_V^2+3g_A^2)F(E_e)\Big[1+a\frac{\vec{p_e}\cdot\vec{p_{\nu}}}{E_eE_{\nu}}+\vec{\sigma_n} \cdot
\big(A\frac{\vec{p_e}}{E_e}+B\frac{\vec{p_{\nu}}}{E_{\nu}} \big)\Big]
\end{equation}

\begin{figure}
\begin{center}
  \includegraphics[width=0.5\textwidth, angle=270]{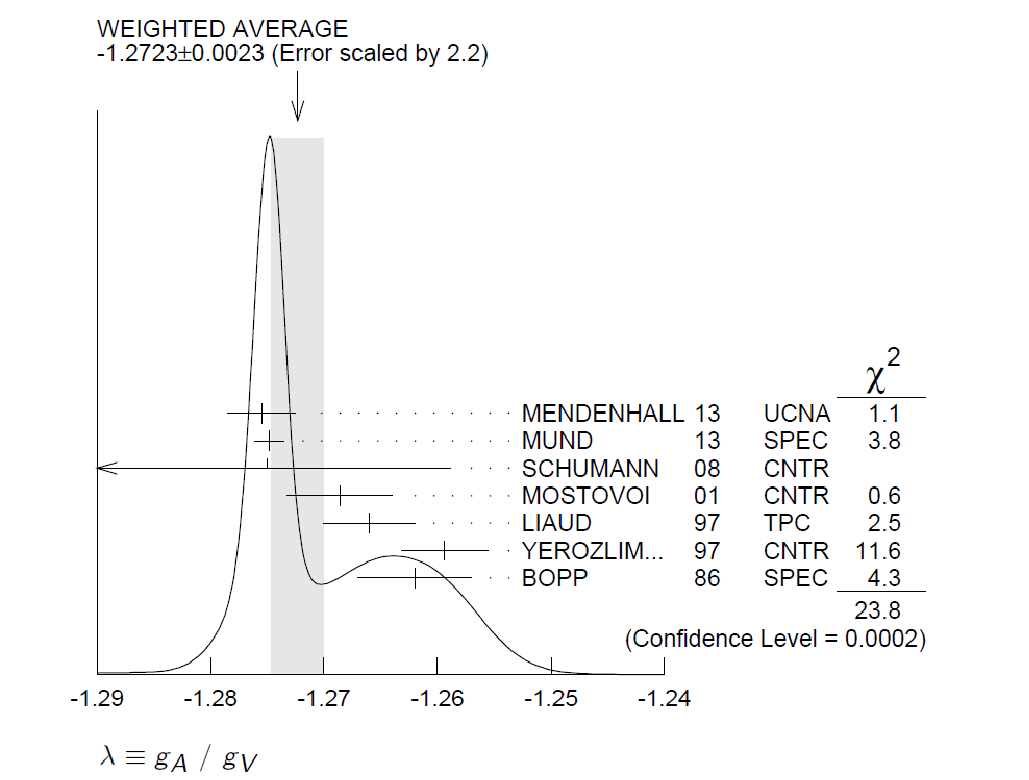} 
\caption{PDG Determination of $\lambda$. Currently used experimental values as well as the average and the scaling factor effect are shown.}
\label{fig:lambda}
\end{center}
\end{figure}

The $(g_V^2+3g_A^2)$ term is proportional to the neutron lifetime, as it is the linear 
combination of the $g_A$ and $g_V$ couplings; whereas the $a, A, B$ correlation coefficients 
can be used to extract $\lambda=\frac{g_A}{g_V}$, their ratio via:
\begin{eqnarray} 
  A=-2\frac{|\lambda|^2+Re[\lambda]}{1+3|\lambda|^2}\quad B=-2\frac{|\lambda|^2-Re[\lambda]}{1+3|\lambda|^2}\quad a=\frac{1-|\lambda|^2}{1+3|\lambda|^2}
\end{eqnarray}
Examining Eq.~\ref{eq:decayrate}, the physical meaning of the correlation parameters is quite clear: 
$A$ is the correlation between the initial spin of the neutron and the momentum of the outgoing electron, 
$B$ is the analog for the anti-neutrino, and $a$ is the correlation between the directions of the 
outgoing electron and anti-neutrino.  The PDG value of $\lambda$ is currently 1.2723$\pm0.0023$~\cite{PDG2016}, 
derived primarily from $A$ measurements.   The final result from the UCNA collaboration~\cite{Brown:2017mhw} 
is consistent with this, but not yet incorporated into the PDG average.   A scaling factor of 2.2 is at present 
assigned to the average uncertainty from the experiments.  This is due to the fact that the most recent and 
precise measurements have shifted towards higher values of $|\lambda|$, resulting in a double peak, 
seen in Fig.~\ref{fig:lambda}. Forthcoming values by way of $a$ measurements from 
aCORN~\cite{few09acorn} and Nab~\cite{Dinko09, Baessler14} experiments will be a necessary 
high-precision check to the existing measurements.
\subsection{Beam Method}
The result of most recent extraction of the neutron lifetime via the ``beam'' method~\cite{Nico05,Yue13} 
is called Beam Lifetime 1 (BL1) and is succeeded by the in-progress measurement, Beam Lifetime 2 (BL2).  These experiments are
done at the National Institute of Standards and Technology in
Gaithersburg, MD (NIST).  This apparatus is one of a kind worldwide and is currently capable of achieving a 1~s
lifetime measurement, which is the level of uncertainty in “storage” experiments.  

Ion-implanted silicon detectors were used during BL1 and are, once again, being used in BL2 for proton detection.  
In the beta decay of cold neutrons (energies of order of 10s of meV), the daughter protons have a maximum of 
760$~$eV of kinetic energy. This is not sufficient for detection, as approximately 10$~$keV is required to penetrate the 
dead layer of the detector. To address this, the proton detector in the experiment is held at a 
-30$~$kV potential, accelerating the particles into the detector.  BL1 counted the protons using a 
discriminator threshold placed on the proton detector signal, meaning the waveform itself was not recorded.  
Protons are sometimes scattered from the detector back into the magnet volume, these particles will be 
guided and accelerated by the EM field back into the detector face.  In this situation the detector will 
see no hits above threshold, or multiple hits from a single particle.  The threshold method of proton counting
will misclassify such events and the waveform data would be unavailable for more detailed analysis.  
The calculation of the backscattering 
correction and uncertainty requires a series of measurements using different acceleration potentials as well as 
detectors with different dead-layer thicknesses.  Essentially, BL1 calculated the neutron lifetime in 
configurations yielding different fractions of protons that backscatter (calculated via SRIM~\cite{SRIM} ) and
extrapolated to zero backscattering. The current uncertainty associated with this process is 0.4~s 
and is unchanged from BL1 to BL2, as no improvements nor changes have been made.

It is likely that the discrepancy between the different determinations of the neutron lifetime are due to an 
unaccounted for (or improperly accounted for) systematic error in one or both approaches.  
The beam method relies on precise and accurate knowledge of the neutron and proton detector efficiencies.  
We propose a method for absolute proton detector calibration through coincidence measurements of electrons and protons from the 
laser-ionization of atomic hydrogen will address the question of hidden ``unknowns'' in the latter. 
Additionally, 60\% of the existing 0.4$~$s uncertainty in the proton backscattering determination 
comes from the calibration of the detector with radioactive sources in-situ and is limited by a 
number of technical issues. A direct calibration will reduce this 0.24$~$s contribution to relatively negligible levels. 

\section{Proposed Method}

The main objective of this publication is to provide a procedure for absolute
detector calibration for both the existing and future proton detector designs. Some collaborations have
built proton sources to attempt to characterize charged particle detectors. Such an approach is
insufficient for lifetime determinations, as the initial flux cannot be measured down to each individual particle. 
Instead, using this method, protons
and electrons would be detected in coincidence (as well as in singles mode) following a laser-induced
ionization of hydrogen gas. 

\begin{figure}[!ht]
\begin{center}
\includegraphics[width=0.45\textwidth]{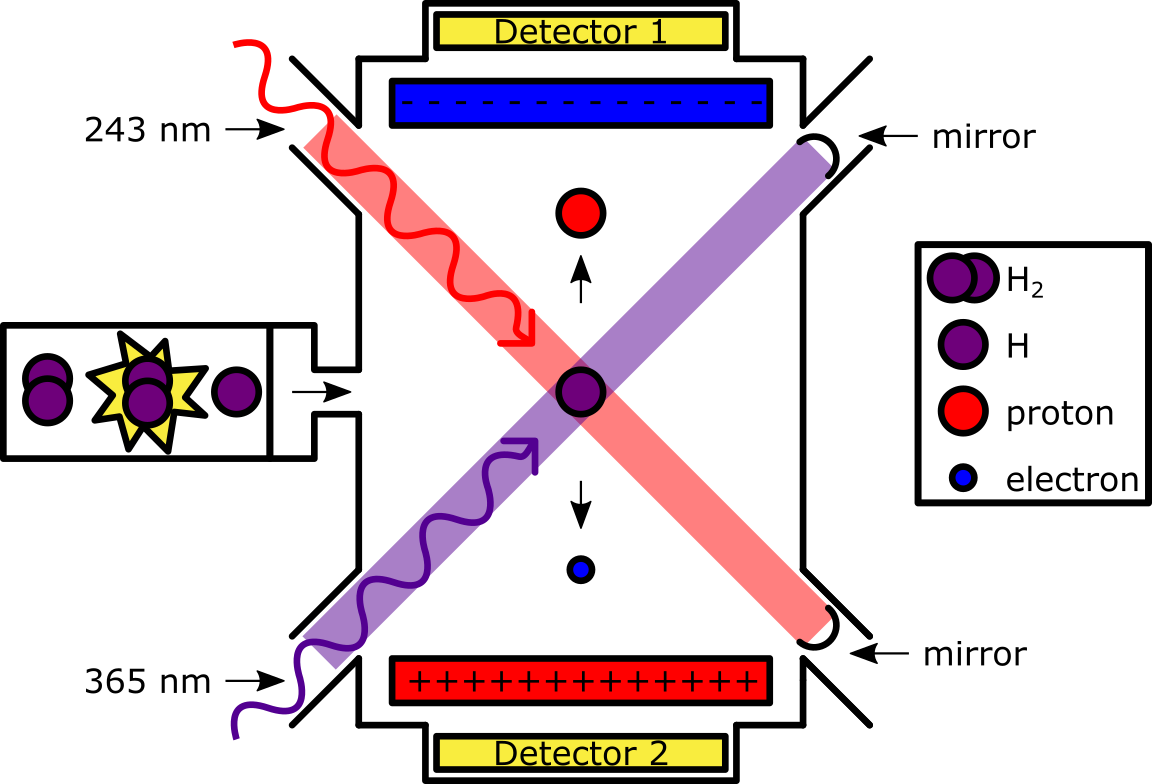}
\includegraphics[width=0.45\textwidth]{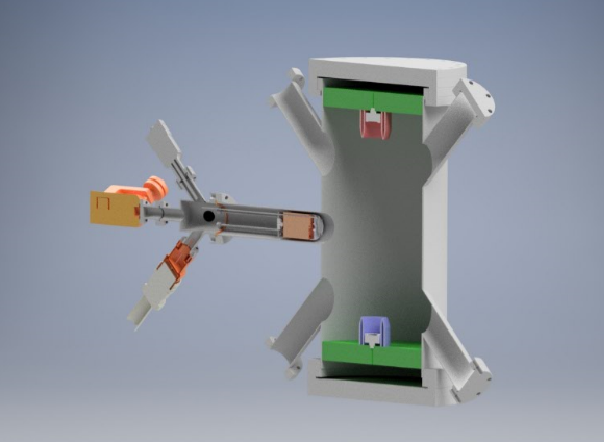}
\caption [width=0.9\textwidth]{Graphic to illustrate the proposed apparatus. Left Panel: molecular hydrogen is disassociated into atomic hydrogen, which is then excited to the $2s$ state via laser excitation.  A higher wavelength laser is then used to ionize the atoms.  The resulting electrons and protons are accelerated by $\pm$ potentials to identical ion-implanted silicon detectors. Right Panel: A notional Inventor model to illustrate the proposed apparatus.  The two arms on the right may also include mirrors to reflect the light for additional efficiency.  The red and blue objects are corona shields to supply the acceleration potentials to the detectors  G10 disks are included for electrical isolation (green), but the HV isolation cages are not pictured.}
\end{center}
\label{fig:calib}
\end{figure}

\subsection{Coincidence e-p detection}
The proposed proton detector calibration apparatus is designed to use singles and 
coincidence detection of electrons and protons from ionization of hydrogen to calibrate 
both detectors (two BL2 detectors will be used). The efficiency of each detector is given as
$\epsilon_1 \& \epsilon_2$, the rate of ionization of hydrogen is $N_0$ and the number
of particles detected in each detector is given by $N_1 \& N_2$.  $N_c$ represents
the number of coincidences detected in the apparatus.  Equation \ref{eq:effs} shows how the
measurements allow for the determination of the individual detector efficiencies.

\begin{equation}
  \label{eq:effs}
    \begin{split}
    N_{1} = N_{0} \epsilon_{1} \quad N_{2} = N_{0} & \epsilon_{2} \quad N_{c} = N_{0} \epsilon_{1} \epsilon_{2} \\
    \epsilon_{1} = \frac{N_c}{N_2} \quad  & \quad  \epsilon_2  = \frac{N_c}{N_1} 
    \end{split}
\end{equation} 

Today, an off-the-shelf atomic hydrogen~\cite{cracker} 
source can be obtained, removing the need to construct something to disassociate 
molecular hydrogen.
Fig.~\ref{fig:calib} shows a schematic of the experimental setup. Atomic hydrogen is 
introduced using a commercially available source (such as the Tectra thermal 
hydrogen cracker~\cite{cracker}). 

\subsection{Laser Excitation Schemes}
Ionizing an H atom requires 13.6 eV of energy.
We propose ionizing the H atoms via a two step laser excitation scheme.
First, we excite the atoms from the $n=1$ ground state to $n=2$ excited state which requires 10.2 eV of energy corresponding to one 122 nm photon or two 243 nm photons.
The one photon transition to the $2p_{3/2}$ state is electric dipole allowed, while a two photon transition to the $2s_{1/2}$ state is required by selection rules \cite{foot}.
Second, the excited atoms are then ionized by a commercially available 365 nm laser \cite{toptica} which provides the remaining 3.4 eV of energy at threshold.
The main technical challenge, recently overcome as will be described, is producing the deep UV excitation laser light.

A common technique to generate the 243 nm photons is frequency doubling 486 nm light using a nonlinear crystal such as beta barium borate (BBO) \cite{foot90} or cesium lithium borate (CLBO) \cite{burkley17}.
The 486 nm light itself can be produced directly by a dye laser \cite{foot90} or alternatively by frequency doubling light at 972.5 nm produced by an external cavity laser diode system \cite{kola06}.
Recently, a frequency quadrupled laser system using Yb-doped fiber amplifiers produced watts of 243 nm light \cite{burkley19}. 
Furthermore, this state of the art laser system was used to populate the 2s state of atomic hydrogen for a precision measurement of the 2s to 8d transition frequency in atomic hydrogen \cite{brandt22}.
Finally, we expect that a frequency quadrupled 972.5 nm 5 W Ti:Sapphire laser would be capable of producing hundreds of mW of 243 nm light \cite{degenkolb}.

Alternatively, two groups have recently demonstrated a pulsed Lyman-$\alpha$ laser at 122 nm \cite{michan15, gabrielse18}. 
The two schemes are similar and involve first the generation of 729 nm using a pulsed Ti:Sapphire laser which is subsequently frequency doubled used a lithium borate (LBO) nonlinear doubling crystal.
The subsequent 364.5 nm light is frequency tripled using third harmonic generation in a Kr/Ar gas cell.
Such a laser was recently used to laser cool antihydrogen at CERN \cite{baker21}.

\subsection{Population Rate Equations}
The fractional populations of the ground state $n_g$, excited state $n_e$, and ionized state $n_i$ vary in location and time.
We approximate the spatial dependence of the various rates by a time dependence assuming that all the H atoms are moving at a constant speed given by the mean thermal speed.
In particular, the laser excitation rate is nonzero only during the time it takes the H atom to traverse the interaction region where the two lasers overlap with the H atomic beam.
All other rates are independent of time although they may be electric field dependent, which we assume is spatially uniform.
The rate equations that govern the populations under these assumptions are:
\begin{eqnarray}
	\dot{n}_g & = & -R n_{g} + \frac{n_e}{\tau_e} \\
	\dot{n}_e & = & +R n_{g} - \frac{n_e}{\tau_e} - \Gamma_\mathrm{ion} n_e \\
	\dot{n}_i & = & + \Gamma_\mathrm{ion} n_e 
\end{eqnarray}
where $n_g + n_e + n_i = 1$, $R$ is the laser excitation rate to the $n=2$ state from the $n=1$ ground state, $\tau_e$ is the lifetime of the excited state in an electric field, and $\Gamma_\mathrm{ion}$ is the photoionization rate from the excited state.
It is straightforward to show that the $E$-field ionization rates from either state as are negligibly small.
If the drift time of the electrons to exit the interaction region is sufficiently short, then the electron-ion recombination rate in the interaction region is also negligibly small.

\subsection{Ion Population Fraction Estimates}
Assuming that the atoms stay in the interaction region long enough for the laser excitation process to reach quasi-equilibrium (i.e. $1/ \tau_e$ is the fastest rate) and that $n_e,n_i \ll n_g \rightarrow n_g \approx 1$, then the ionized population fraction can be estimated by:
\begin{equation}
n_i \approx R \tau_e \Gamma_\mathrm{ion} t_\mathrm{int}
\end{equation}  
where $t_\mathrm{int}$ is the effective time that atoms are in the interaction region.
We assume the following parameters to estimate the rates and ion population fractions under two different laser excitation schemes:
\begin{itemize}
\item $P_{2\gamma} = 1\ \mathrm{W}$, CW power of the 243 nm two photon excitation laser \cite{burkley19}
\item $w_{2\gamma} = 100\ \mu\mathrm{m}$, gaussian beam radius for the 243 nm laser 
\item $E_\mathrm{pulse} = 1\ \mathrm{nJ}$, $\tau_\mathrm{pulse} = 15\ \mathrm{ns}$, $\nu_\mathrm{rep} = 10\ \mathrm{Hz}$, pulse energy, duration, and repetition rate for a 122 nm single photon exception laser \cite{michan15,gabrielse18}
\item $w_{1\gamma} = 1\ \mathrm{mm}$, gaussian beam radius for the 122 nm laser \cite{baker21}
\item $P_\mathrm{ion} = 1\ \mathrm{W}$, CW power of the 365 nm photoionization laser \cite{toptica}
\item $w_{ion} = 100\ \mu\mathrm{m}$, gaussian beam radius for the 365 nm laser 
\item $\ell_\mathrm{int} = 2\ \mathrm{mm}$, length of interaction region
\item $T = 2000\ \mathrm{K}$, temperature of H atomic beam \cite{bb93,elw98}
\item $E_\mathrm{field} = 3\ \mathrm{kV/cm}$
\end{itemize}
Assuming a gaussian laser beam and a Doppler-broadened atomic beam, the peak laser excitation rates can be estimated by \cite{siegman}:
\begin{eqnarray}
R_{2\gamma} & = & \left ( \frac{2P_{2\gamma}}{\pi w_{2\gamma}^2 h \nu_{2\gamma}} \right ) \left [  \left (\frac{\sigma_{2\gamma}}{g(2\omega)I(\omega)} \right )  \sqrt{ \frac{m_e c^2}{2\pi kT \omega^2_{2\gamma}} }  \left ( \frac{2P_{2\gamma}}{\pi w_{2\gamma}^2} \right ) \right ]  \\
& = & \left ( 121\ \mathrm{\mu Hz} \right ) \left [ \frac{P_{2 \gamma}}{1\ \mathrm{W}} \right ]^2 \left [ \frac{100\ \mathrm{\mu m}} {w_{2\gamma}} \right ]^4  \sqrt{\frac{2000\ \mathrm{K}}{T}} \\
R_{1\gamma} & = & \left ( \frac{2E_\mathrm{pulse}/\tau_\mathrm{pulse}}{\pi w_{1\gamma}^2 h \nu_{1\gamma}} \right ) \left [  \pi r_e c f_{2p3/2}  \sqrt{ \frac{m_e c^2}{2\pi kT \nu^2_{1\gamma}} }  \right ]  \\
& = & \left ( 5.33\ \mathrm{kHz} \right ) \left [ \frac{E_\mathrm{pulse}/\tau_\mathrm{pulse}}{(1\ \mathrm{nJ})/(15\ \mathrm{ns})} \right ] \left [ \frac{1\ \mathrm{m m}} {w_{1\gamma}} \right ]^2 \sqrt{\frac{2000\ \mathrm{K}}{T}} \\
\Gamma^{2s,2p}_\mathrm{ion} & = & \left ( \frac{2P_\mathrm{ion}}{\pi w_\mathrm{ion}^2 h \nu_\mathrm{ion}} \right ) \left [  \sigma^{2s,2p}_\mathrm{ion}(0)  \right ]  \\
& =& \left [ \begin{array}{c} (173\ \mathrm{kHz})_{2s} \\ (158\ \mathrm{kHz})_{2p} \end{array} \right ]  \left [ \frac{P_\mathrm{ion}}{1\ \mathrm{W}} \right ] \left [ \frac{100\ \mathrm{\mu m}} {w_\mathrm{ion}} \right ]^2
\end{eqnarray}
where the term in $\left (\cdots \right )$ is the photon flux (number of photons per unit time per unit area) and the term in $\left [ \cdots \right ]$ is the laser interaction cross section.
The full Stark-mixed excited state lifetimes for $E_\mathrm{field} = 3\ \mathrm{kV/cm}$ are $\tau_{2s} = \tau_{2p} = 3.19\ \mathrm{ns}$.
The interaction time is given by time needed for the atomic beam to traverse the interaction region for the two photon scheme $\tau_\mathrm{int}(2\ \mathrm{mm},2000\ \mathrm{K}) = \ell_\mathrm{int} \sqrt{\pi m_H/(8kT)} = 0.309\ \mathrm{\mu s}$ or the pulse duration for the one photon scheme $\tau_\mathrm{int} = \tau_\mathrm{pulse} = 15\ \mathrm{ns}$.

Assuming a thermal atomic beam temperature of $T = 2000\ \mathrm{K}$ and a static electric field of $E_\mathrm{field} =  3\ \mathrm{kV/cm}$, the estimated ionization fraction for the two laser excitation schemes are $n^{2\gamma}_i = \sci{2}{-14}$ and $n^{1\gamma}_i = \sci{4}{-8}$. 
The 122 nm excitation laser scheme, although technically more challenging, provides a $10^6$ higher ionization fraction due to the significantly higher photoexcitation cross section.

\subsection{Ion Production Rate Estimates}
The ion production rate can be estimated by $dN_i/dt = n_i dN/dt$ where $dN/dt$ is the number of H atoms passing through the interaction region.
A commercial H atomic beam source (HABS) produces a total output of $\sci{2}{15}$ atoms/s with a central peak flux of $\sci{1.5}{13} \ \mathrm{atoms/cm^2/s}$ with a thermal velocity distribution corresponding to $T = 2000\ \mathrm{K}$ \cite{bb93,elw98,cracker}.
Assuming a intersection cross sectional area of $\pi w_\mathrm{ion}^2 = \sci{3}{-4}\ \mathrm{cm^2} $ and using this hot commercial HABS with $dN/dt = \sci{5}{9}$ atoms/s would result in about $dN_i^{2\gamma}/dt \approx 10$ ions per day for the two photon scheme and  $dN_i^{1\gamma}/dt \approx 200$ ions per sec for the one photon scheme.

Recently a cryogenic HABS has been developed \cite{cooper20} and used for a precision measurement of the 2s to 8d transition frequency in atomic hydrogen \cite{brandt22}.
This cryogenic source produces a total output of $10^{17}$ H atoms/s at $T = 6\ \mathrm{K}$.
Assuming a similar angular distribution as the hot commercial HABS, we estimate $dN/dt = \sci{2}{11}$ atoms/s which would result in about $dN_i^{2\gamma}/dt \approx 2$ ions per sec for the two photon scheme and  $dN_i^{1\gamma}/dt \approx 200k$ ions per sec for the one photon scheme.
The increased ion production rates are due to both the increased atomic beam flux and the significantly smaller Doppler broadening at cryogenic temperatures. 

In the case that a background rate is proving problematic, the light sources may be operated in pulsed mode.  This would help by allowing the light source trigger to act as a time-zero for the cracking of the hydrogen.  The electron and proton measurement windows could then be opened at a well defined time after t=0.

\begin{figure}[!ht] 
  \begin{center}
    \includegraphics[width=0.43\textwidth]{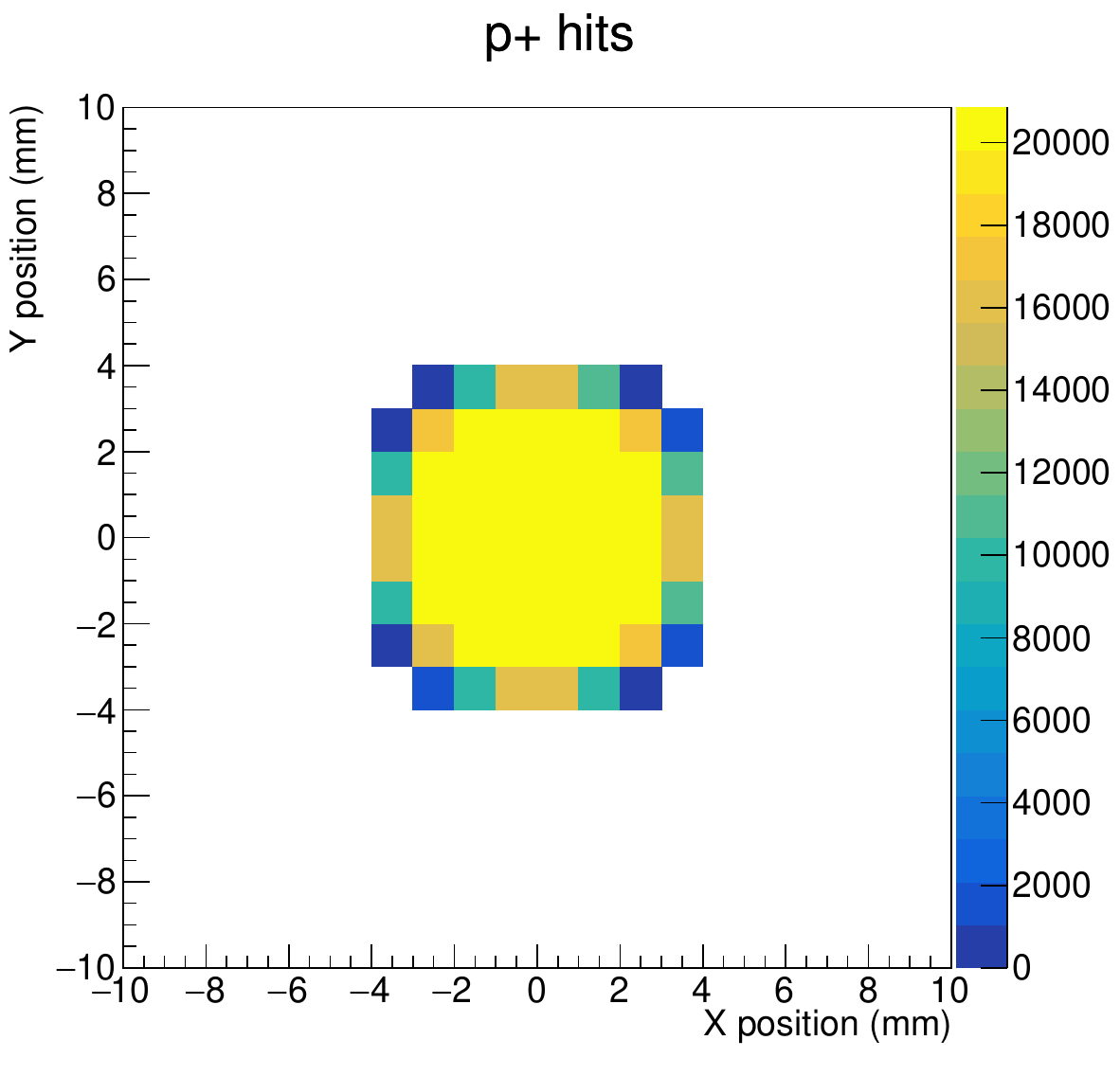}
    \includegraphics[width=0.43\textwidth]{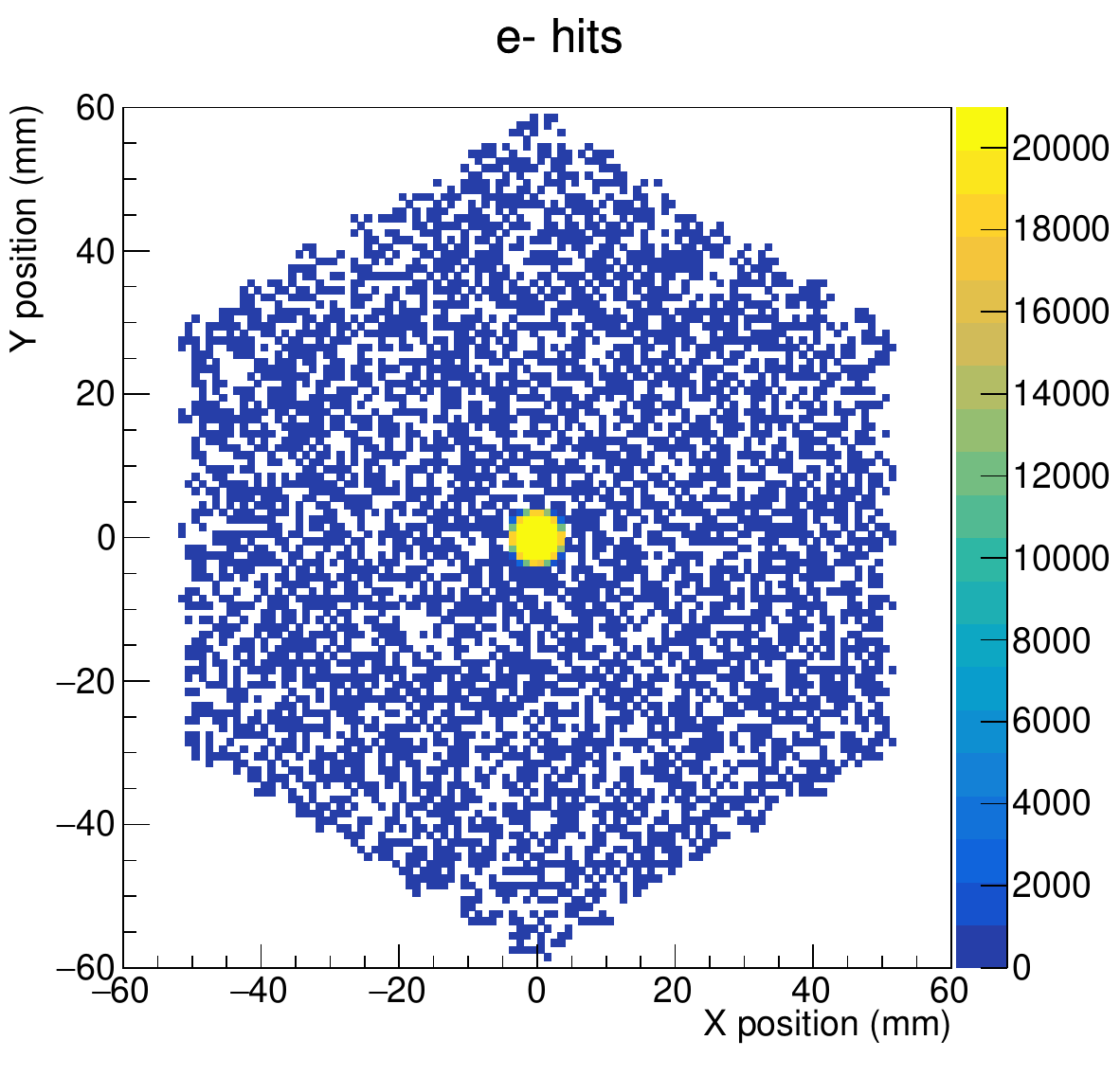}
    \caption [width=0.9\textwidth]{Preliminary results from a Geant4 simulation of proton and electron transport. 
    Left Panel: A density plot showing the area of proton hits on the detector.  The centroid will 
    be offset in x by a few mm in the real device due to the initial momentum of hydrogen exiting the cracker.  
    Right Panel: A density plot showing the area of electron hits on the detector as simulated in Geant4. The initial hit is always in the 
    yellow central area.  The wider area of hits are secondary and tertiary electron back-scatters from the detector face.  Both the proton 
    hit area and the initial hit area for electrons are well within the smallest detector size the apparatus is expected to test (19.6 mm diameter).}
\label{fig:hits}
\end{center}
\end{figure} 

 The electrons and protons will be accelerated using a $\pm$30kV potential to the detectors on 
 either side of the vacuum chamber.   This is essentially replicating the proton detection arm 
 of the BL1/2 apparatus twice (for positive and negative charge detection).   In the lifetime 
 experiment itself, the protons are guided to the detector using the magnetic field, which is 
 not present in this proposed apparatus.  
 The simulation shown in Fig.~\ref{fig:hits} creates a proton-electron pair at the center of the apparatus from a hydrogen atom whose velocity is 
 6000~$m/s$ (determined by the temperature of the hydrogen cracker).  The particles see a 60~kV electric 
 field across a distance of 20~cm between the detectors. 
 Fig.~\ref{fig:hits} shows the area of the proton and electron hits.  
 The smallest detectors used in BL2 have a diameter of 19.6~mm, meaning the initial hits of particles should not 
 fall outside the active area of the detector.  
 The silicon detectors will scatter a fraction of incoming charged particles back into the drift area after absorbing some initial energy deposit.  
 The back-scattered particles will decelerate into the potential region and be re-accelerated into the detector.  It is shown in Fig.~\ref{fig:hits}
 that the electrons frequently scatter laterally as well, causing the subsequent hits to increase in radius from the center of the detector.  The Protons
 do not show this behavior.  The detector interaction is handled using the Livermore low energy physics package \cite{livermore} included in Geant4.

\begin{figure}[!ht] 
  \begin{center}
    \includegraphics[width=0.43\textwidth]{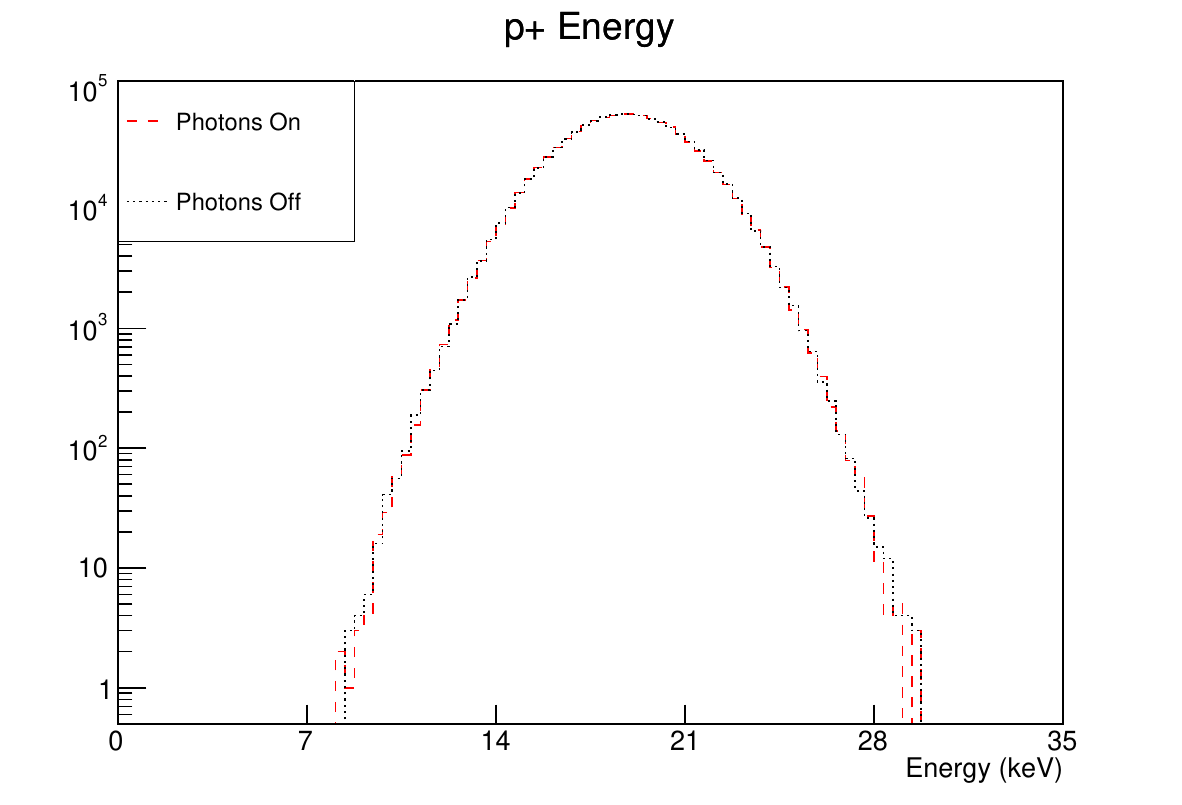}
    \includegraphics[width=0.43\textwidth]{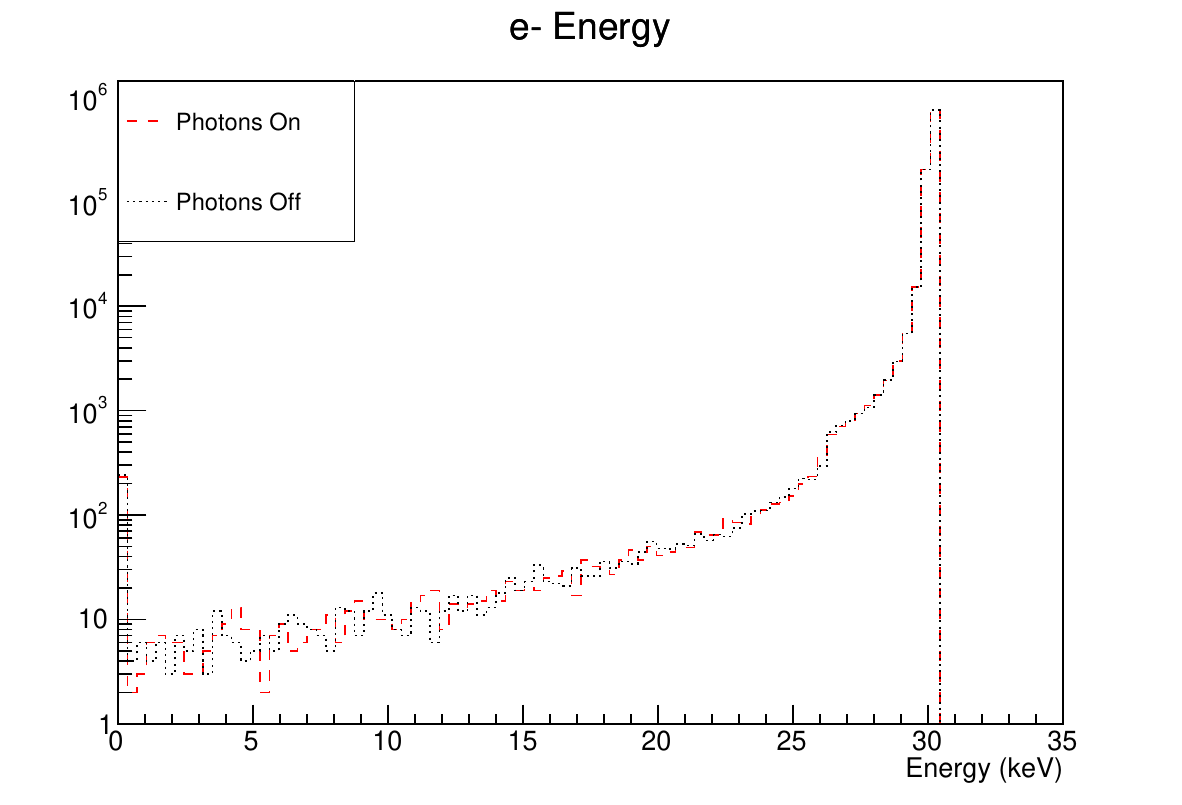}
    \caption [width=0.9\textwidth]{Preliminary results from a Geant4
    simulation of proton and electron transport. 
    Protons and electrons were thrown with coincident optical photons to simulate the chamber being illuminated by the wavelengths necessary to
    split the atomic hydrogen.  No significant contribution to the number of hits nor the hit energy spectra is observed when optical photons are
    included in the simulation (dashed line) compared to when they are not (dotted line).}
\label{fig:opph}
\end{center}
\end{figure} 

 Additional Geant4 simulations were performed, throwing optical photons at the detectors using the wavelengths of interest.  This simulation
 was designed to test the device in a running mode, with protons, electrons and photons being thrown simultaneously.  The study
 indicates that photons do not cause additional energetic deposits in the detector material. Shown in Fig.~\ref{fig:opph} is an energy spectrum of
 proton and electron hits from two different Geant4 simulations, one includes optical photons in the particle gun the other does not.  
 No significant differences 
 in total number of hits, or the energy distribution of hits is observed.  The optical photon gun was tested in two configurations, positioned in the 
 center of the device pointed normal to the detector face, and positioned 2mm from the detector with an incident angle on the silicon of $30\deg$. 
 Neither of these scenarios produced noticeable differences in the hit count nor energy spectrum of electron and proton hits. A highly
 absorptive coating will be painted on the inside of the apparatus to minimize the bouncing of optical photons around the chamber, 
 in the true setup the lights are also not pointed directly at the detectors.  This simulation was meant to represent a pessimistic case
 where neither of those factors were included.

\section{Summary}

The method proposed above would constitute vital part of the fundamental neutron physics program. To reach
the precision where neutron measurements can be used to test the Standard Model require resolving the
discrepancy in the neutron lifetime results from different measurement methods. The mounting of future
experiments with low goal uncertainties demand that systematic effects are well understood. Accurate
proton detection and absolute detector calibration is a crucial part of this future.
\include{jts-comments}
\clearpage

\bibliography{mybibfile}

\begin{thebibliography}{10}
\expandafter\ifx\csname url\endcsname\relax
  \def\url#1{\texttt{#1}}\fi
\expandafter\ifx\csname urlprefix\endcsname\relax\def\urlprefix{URL }\fi
\expandafter\ifx\csname href\endcsname\relax
  \def\href#1#2{#2} \def\path#1{#1}\fi

\bibitem{Wietfeldt11}
F.~E. Wietfeldt, G.~L. Greene, The neutron lifetime, Rev. Mod. Phys. 83 (2011)
  1173--1192.

\bibitem{Abele2008}
H.~Abele, The neutron. its properties and basic interactions, Prog. Part. Nucl.
  Phys. 60 (2008) 1 -- 81.

\bibitem{Planck}
P.~A.~R. Ade, et~al., {Planck 2013 results. XVI. Cosmological parameters},
  Astron. Astrophys. 571 (2014) A16.
\newblock \href {http://arxiv.org/abs/1303.5076} {\path{arXiv:1303.5076}},
  \href {https://doi.org/10.1051/0004-6361/201321591}
  {\path{doi:10.1051/0004-6361/201321591}}.

\bibitem{iocco09}
F.~{Iocco}, G.~{Mangano}, G.~{Miele}, O.~{Pisanti}, P.~D. {Serpico},
  {Primordial nucleosynthesis: From precision cosmology to fundamental
  physics}, Phys.~Rep. 472 (2009) 1--76.
\newblock \href {https://doi.org/10.1016/j.physrep.2009.02.002}
  {\path{doi:10.1016/j.physrep.2009.02.002}}.

\bibitem{nollett14}
K.~M. Nollett, G.~Steigman,
  \href{http://link.aps.org/doi/10.1103/PhysRevD.89.083508}{Bbn and the cmb
  constrain light, electromagnetically coupled wimps}, Phys. Rev. D 89 (2014)
  083508.
\newblock \href {https://doi.org/10.1103/PhysRevD.89.083508}
  {\path{doi:10.1103/PhysRevD.89.083508}}.
\newline\urlprefix\url{http://link.aps.org/doi/10.1103/PhysRevD.89.083508}

\bibitem{gonzalez}
F.~M. Gonzalez, E.~M. Fries, C.~Cude-Woods, T.~Bailey, M.~Blatnik, L.~J.
  Broussard, N.~B. Callahan, J.~H. Choi, S.~M. Clayton, S.~A. Currie, M.~Dawid,
  E.~B. Dees, B.~W. Filippone, W.~Fox, P.~Geltenbort, E.~George, L.~Hayen,
  K.~P. Hickerson, M.~A. Hoffbauer, K.~Hoffman, A.~T. Holley, T.~M. Ito,
  A.~Komives, C.-Y. Liu, M.~Makela, C.~L. Morris, R.~Musedinovic,
  C.~O'Shaughnessy, R.~W. Pattie, J.~Ramsey, D.~J. Salvat, A.~Saunders, E.~I.
  Sharapov, S.~Slutsky, V.~Su, X.~Sun, C.~Swank, Z.~Tang, W.~Uhrich,
  J.~Vanderwerp, P.~Walstrom, Z.~Wang, W.~Wei, A.~R. Young,
  \href{https://link.aps.org/doi/10.1103/PhysRevLett.127.162501}{Improved
  neutron lifetime measurement with $\mathrm{UCN}\ensuremath{\tau}$}, Phys.
  Rev. Lett. 127 (2021) 162501.
\newblock \href {https://doi.org/10.1103/PhysRevLett.127.162501}
  {\path{doi:10.1103/PhysRevLett.127.162501}}.
\newline\urlprefix\url{https://link.aps.org/doi/10.1103/PhysRevLett.127.162501}

\bibitem{Hardy09}
J.~C. Hardy, I.~S. Towner, Superallowed ${0}^{+} \longrightarrow {0}^{+}$
  nuclear beta decays: A new survey with precision tests of the conserved
  vector current hypothesis and the standard model, Physical Review C 79 (2009)
  055502.

\bibitem{Hardy2014}
J.~C. Hardy, I.~S. Towner, {Superallowed $0^+\to 0^+$ nuclear $\beta$ decays:
  2014 critical survey, with precise results for $V_{ud}$ and CKM unitarity},
  Phys. Rev. C91~(2) (2015) 025501.
\newblock \href {http://arxiv.org/abs/1411.5987} {\path{arXiv:1411.5987}},
  \href {https://doi.org/10.1103/PhysRevC.91.025501}
  {\path{doi:10.1103/PhysRevC.91.025501}}.

\bibitem{Jackson57}
J.~D. Jackson, S.~B. Treiman, H.~W.~W. Jr, Possible tests of time reversal
  invariance in beta decay, Physical Review 106 (1957) 517--521.

\bibitem{PDG2016}
C.~Patrignani, et~al., {Review of Particle Physics}, Chin. Phys. C40~(10)
  (2016) 100001.

\bibitem{Brown:2017mhw}
M.~A.~P. Brown, et~al., {New result for the neutron $\beta$-asymmetry parameter
  $A_0$ from UCNA} (2017).
\newblock \href {http://arxiv.org/abs/1712.00884} {\path{arXiv:1712.00884}}.

\bibitem{few09acorn}
F.~Wietfeldt, et~al., a{CORN}: An experiment to measure the electron -
  antineutrino correlation in neutron decay, Nucl. Instr. Meth. Phys. Res. A
  611 (2009) 207 -- 210.
\newblock \href {https://doi.org/10.1016/j.nima.2009.07.064}
  {\path{doi:10.1016/j.nima.2009.07.064}}.

\bibitem{Dinko09}
D.~Po{\v c}ani{\' c}, et~al., Nab: Measurement principles, apparatus and
  uncertainties, Nucl. Instr. Meth. Phys. Res. A 611 (2009) 211 -- 215.
\newblock \href {https://doi.org/10.1016/j.nima.2009.07.065}
  {\path{doi:10.1016/j.nima.2009.07.065}}.

\bibitem{Baessler14}
S.~Bae{\ss}ler, J.~D. Bowman, S.~Penttil{\"a}, D.~Po{\v c}ani{\' c}, New
  precision measurements of free neutron beta decay with cold neutrons, J.
  Phys. G 41~(11) (2014) 114003.
\newblock \href {https://doi.org/10.1088/0954-3899/41/11/114003}
  {\path{doi:10.1088/0954-3899/41/11/114003}}.

\bibitem{Nico05}
J.~S. Nico, et~al., Measurement of the neutron lifetime by counting trapped
  protons in a cold neutron beam, Phys. Rev. C 71 (2005) 055502.

\bibitem{Yue13}
A.~Yue, M.~Dewey, D.~Gilliam, G.~Greene, A.~Laptev, et~al., {Improved
  Determination of the Neutron Lifetime}, Phys.Rev.Lett. 111~(22) (2013)
  222501.
\newblock \href {http://arxiv.org/abs/1309.2623} {\path{arXiv:1309.2623}},
  \href {https://doi.org/10.1103/PhysRevLett.111.222501}
  {\path{doi:10.1103/PhysRevLett.111.222501}}.

\bibitem{SRIM}
J.~F. Ziegler, Stopping range of ions in matter, computer code available at
  www.srim.org (2008).

\bibitem{cracker}
Tectra, Atomic hydrogen source,
  http://tectra.de/sample-preparation/atomic-hydrogen-source/.

\bibitem{foot}
C.~J. Foot, Atomic physics, Oxford University Press, Oxford, UK, 2005.

\bibitem{toptica}
{TOPTICA Photonics AG},
  \href{https://www.toptica.com/products/single-frequency-lasers/uv-rgb-solutions}{Products:uv\/rgb
  solutions}, [Online; accessed 31-October-2023] (2023).
\newline\urlprefix\url{https://www.toptica.com/products/single-frequency-lasers/uv-rgb-solutions}

\bibitem{foot90}
C.~J. Foot, P.~Hannaford, D.~N. Stacey, C.~D. Thompson, G.~H. Woodman, P.~E.~G.
  Baird, J.~B. Swan, G.~K. Woodgate,
  \href{https://dx.doi.org/10.1088/0953-4075/23/11/003}{Observation of the
  1s-2s transition in atomic hydrogen in an atomic beam}, Journal of Physics B:
  Atomic, Molecular and Optical Physics 23~(11) (1990) L203.
\newblock \href {https://doi.org/10.1088/0953-4075/23/11/003}
  {\path{doi:10.1088/0953-4075/23/11/003}}.
\newline\urlprefix\url{https://dx.doi.org/10.1088/0953-4075/23/11/003}

\bibitem{burkley17}
Z.~Burkley, C.~Rasor, S.~F. Cooper, A.~D. Brandt, D.~C. Yost,
  \href{https://doi.org/10.1007/s00340-016-6583-9}{Yb fiber amplifier at 972.5
  nm with frequency quadrupling to 243.1 nm}, Applied Physics B 123~(1) (2016)
  5.
\newblock \href {https://doi.org/10.1007/s00340-016-6583-9}
  {\path{doi:10.1007/s00340-016-6583-9}}.
\newline\urlprefix\url{https://doi.org/10.1007/s00340-016-6583-9}

\bibitem{kola06}
N.~Kolachevsky, J.~Alnis, S.~D. Bergeson, T.~W. H\"ansch,
  \href{https://link.aps.org/doi/10.1103/PhysRevA.73.021801}{Compact
  solid-state laser source for $1s\text{\ensuremath{-}}2s$ spectroscopy in
  atomic hydrogen}, Phys. Rev. A 73 (2006) 021801.
\newblock \href {https://doi.org/10.1103/PhysRevA.73.021801}
  {\path{doi:10.1103/PhysRevA.73.021801}}.
\newline\urlprefix\url{https://link.aps.org/doi/10.1103/PhysRevA.73.021801}

\bibitem{burkley19}
Z.~Burkley, A.~D. Brandt, C.~Rasor, S.~F. Cooper, D.~C. Yost,
  \href{https://opg.optica.org/ao/abstract.cfm?URI=ao-58-7-1657}{Highly
  coherent, watt-level deep-uv radiation via a frequency-quadrupled yb-fiber
  laser system}, Appl. Opt. 58~(7) (2019) 1657--1661.
\newblock \href {https://doi.org/10.1364/AO.58.001657}
  {\path{doi:10.1364/AO.58.001657}}.
\newline\urlprefix\url{https://opg.optica.org/ao/abstract.cfm?URI=ao-58-7-1657}

\bibitem{brandt22}
A.~D. Brandt, S.~F. Cooper, C.~Rasor, Z.~Burkley, A.~Matveev, D.~C. Yost,
  \href{https://link.aps.org/doi/10.1103/PhysRevLett.128.023001}{Measurement of
  the $2{\mathrm{s}}_{1/2}\ensuremath{-}8{\mathrm{d}}_{5/2}$ transition in
  hydrogen}, Phys. Rev. Lett. 128 (2022) 023001.
\newblock \href {https://doi.org/10.1103/PhysRevLett.128.023001}
  {\path{doi:10.1103/PhysRevLett.128.023001}}.
\newline\urlprefix\url{https://link.aps.org/doi/10.1103/PhysRevLett.128.023001}

\bibitem{degenkolb}
S.~Degenkolb, Optical magnetometry using multiphoton transitions, Ph.D. thesis,
  University of Michigan, Ann Arbor (2016).

\bibitem{michan15}
J.~M. Michan, G.~Polovy, K.~W. Madison, M.~C. Fujiwara, T.~Momose,
  \href{https://link.springer.com/article/10.1007/s10751-015-1186-0}{Narrowband
  solid state vuv coherent source for laser cooling of antihydrogen}, Hyperfine
  Interactions~(235) (2015) 29--36.
\newblock \href {https://doi.org/10.1007/s10751-015-1186-0}
  {\path{doi:10.1007/s10751-015-1186-0}}.
\newline\urlprefix\url{https://link.springer.com/article/10.1007/s10751-015-1186-0}

\bibitem{gabrielse18}
G.~Gabrielse, B.~Glowacz, D.~Grzonka, C.~D. Hamley, E.~A. Hessels, N.~Jones,
  G.~Khatri, S.~A. Lee, C.~Meisenhelder, T.~Morrison, E.~Nottet, C.~Rasor,
  S.~Ronald, T.~Skinner, C.~H. Storry, E.~Tardiff, D.~Yost, D.~M. Zambrano,
  M.~Zielinski,
  \href{https://opg.optica.org/ol/abstract.cfm?URI=ol-43-12-2905}{Lyman-$\alpha$
  source for laser cooling antihydrogen}, Opt. Lett. 43~(12) (2018) 2905--2908.
\newblock \href {https://doi.org/10.1364/OL.43.002905}
  {\path{doi:10.1364/OL.43.002905}}.
\newline\urlprefix\url{https://opg.optica.org/ol/abstract.cfm?URI=ol-43-12-2905}

\bibitem{baker21}
C.~Baker, et~al.,
  \href{https://www.nature.com/articles/s41586-021-03289-6}{Laser cooling of
  antihydrogen atoms}, Nature~(592) (2021) 35--42.
\newblock \href {https://doi.org/10.1038/s41586-021-03289-6}
  {\path{doi:10.1038/s41586-021-03289-6}}.
\newline\urlprefix\url{https://www.nature.com/articles/s41586-021-03289-6}

\bibitem{bb93}
U.~Bischler, E.~Bertel, \href{https://doi.org/10.1116/1.578754}{{Simple source
  of atomic hydrogen for ultrahigh vacuum applications}}, Journal of Vacuum
  Science \& Technology A 11~(2) (1993) 458--460.
\newblock \href
  {http://arxiv.org/abs/https://pubs.aip.org/avs/jva/article-pdf/11/2/458/11853287/458\_1\_online.pdf}
  {\path{arXiv:https://pubs.aip.org/avs/jva/article-pdf/11/2/458/11853287/458\_1\_online.pdf}},
  \href {https://doi.org/10.1116/1.578754} {\path{doi:10.1116/1.578754}}.
\newline\urlprefix\url{https://doi.org/10.1116/1.578754}

\bibitem{elw98}
C.~Eibl, G.~Lackner, A.~Winkler,
  \href{https://doi.org/10.1116/1.581449}{{Quantitative characterization of a
  highly effective atomic hydrogen doser}}, Journal of Vacuum Science \&
  Technology A 16~(5) (1998) 2979--2989.
\newblock \href
  {http://arxiv.org/abs/https://pubs.aip.org/avs/jva/article-pdf/16/5/2979/7432502/2979\_1\_online.pdf}
  {\path{arXiv:https://pubs.aip.org/avs/jva/article-pdf/16/5/2979/7432502/2979\_1\_online.pdf}},
  \href {https://doi.org/10.1116/1.581449} {\path{doi:10.1116/1.581449}}.
\newline\urlprefix\url{https://doi.org/10.1116/1.581449}

\bibitem{siegman}
A.~E. Siegman, Lasers, revised edition, University Science Books, Melville, NY,
  USA, 1986.

\bibitem{cooper20}
S.~F. Cooper, A.~D. Brandt, C.~Rasor, Z.~Burkley, D.~C. Yost,
  \href{https://doi.org/10.1063/1.5129156}{{Cryogenic atomic hydrogen beam
  apparatus with velocity characterization}}, Review of Scientific Instruments
  91~(1) (2020) 013201.
\newblock \href
  {http://arxiv.org/abs/https://pubs.aip.org/aip/rsi/article-pdf/doi/10.1063/1.5129156/13992302/013201\_1\_online.pdf}
  {\path{arXiv:https://pubs.aip.org/aip/rsi/article-pdf/doi/10.1063/1.5129156/13992302/013201\_1\_online.pdf}},
  \href {https://doi.org/10.1063/1.5129156} {\path{doi:10.1063/1.5129156}}.
\newline\urlprefix\url{https://doi.org/10.1063/1.5129156}

\bibitem{livermore}
L.~N. Lab, Livermore low energy electromagnetic physics, code Documentation
  available: https://geant4.web.cern.ch/node/1619 (2003).

\bibitem{tung86}
J.~H. Tung, A.~Z. Tang, G.~J. Salamo, F.~T. Chan,
  \href{https://opg.optica.org/josab/abstract.cfm?URI=josab-3-6-837}{Two-photon
  absorption of atomic hydrogen from two light beams}, J. Opt. Soc. Am. B 3~(6)
  (1986) 837--848.
\newblock \href {https://doi.org/10.1364/JOSAB.3.000837}
  {\path{doi:10.1364/JOSAB.3.000837}}.
\newline\urlprefix\url{https://opg.optica.org/josab/abstract.cfm?URI=josab-3-6-837}

\bibitem{burgess65}
A.~Burgess, {Tables of hydrogenic photoionization cross-sections and
  recombination coefficients}, Memoirs of the Royal Astronomical Society 69
  (1965) 1.

\bibitem{LL65}
L.~D. Landau, E.~M. Lifshitz, Quantum Mechanics, Non-relativistic Theory,
  Second edition, revised and enlarged, Pergammon Press, Oxford, U.K., 1965.

\bibitem{rg62}
M.~H. Rice, R.~H. Good,
  \href{https://opg.optica.org/abstract.cfm?URI=josa-52-3-239}{Stark effect in
  hydrogen$\ast$}, J. Opt. Soc. Am. 52~(3) (1962) 239--246.
\newblock \href {https://doi.org/10.1364/JOSA.52.000239}
  {\path{doi:10.1364/JOSA.52.000239}}.
\newline\urlprefix\url{https://opg.optica.org/abstract.cfm?URI=josa-52-3-239}

\bibitem{dk78}
R.~J. Damburg, V.~V. Kolosov,
  \href{https://dx.doi.org/10.1088/0022-3700/11/11/009}{An asymptotic approach
  to the stark effect for the hydrogen atom}, Journal of Physics B: Atomic and
  Molecular Physics 11~(11) (1978) 1921.
\newblock \href {https://doi.org/10.1088/0022-3700/11/11/009}
  {\path{doi:10.1088/0022-3700/11/11/009}}.
\newline\urlprefix\url{https://dx.doi.org/10.1088/0022-3700/11/11/009}

\bibitem{sg51}
J.~{Spitzer}, Lyman, J.~L. {Greenstein}, {Continuous Emission from Planetary
  Nebulae}, Astrophysical Journal 114 (1951) 407.
\newblock \href {https://doi.org/10.1086/145480} {\path{doi:10.1086/145480}}.

\bibitem{nist-asd}
A.~Kramida, {Yu.~Ralchenko}, J.~Reader, {and NIST ASD Team}, {NIST Atomic
  Spectra Database (ver. 5.10), [Online]. Available:
  {\tt{https://physics.nist.gov/asd}} [2023, October 30]. National Institute of
  Standards and Technology, Gaithersburg, MD.} (2022).

\bibitem{bt40}
G.~{Breit}, E.~{Teller}, {Metastability of Hydrogen and Helium Levels.},
  Astrophysical Journal 91 (1940) 215.
\newblock \href {https://doi.org/10.1086/144158} {\path{doi:10.1086/144158}}.

\bibitem{lr50}
W.~E. Lamb, R.~C. Retherford,
  \href{https://link.aps.org/doi/10.1103/PhysRev.79.549}{Fine structure of the
  hydrogen atom. part i}, Phys. Rev. 79 (1950) 549--572.
\newblock \href {https://doi.org/10.1103/PhysRev.79.549}
  {\path{doi:10.1103/PhysRev.79.549}}.
\newline\urlprefix\url{https://link.aps.org/doi/10.1103/PhysRev.79.549}

\end{thebibliography}

\end{document}